\documentclass[journal,12pt,draftclsnofoot,onecolumn]{IEEEtran}
\pdfoutput=1
\newcommand{\href}[1]{}
\usepackage[colorlinks=true,pdfstartview=FitV,linkcolor=black,citecolor=black,urlcolor=blue,plainpages=false]{hyperref}

\usepackage[numbers,sort&compress]{natbib}
\usepackage{hypernat}

\usepackage[pdftex]{graphicx}

\graphicspath{./figs}

\DeclareGraphicsExtensions{.pdf,.jpeg,.png}
 
\usepackage[cmex10]{amsmath}

\usepackage{array}
\usepackage{mdwmath}
\usepackage{mdwtab}

\usepackage[caption=false,font=footnotesize]{subfig}

\usepackage{fixltx2e}

\begin{document}
\title{Building a Cooperative Communications System}

\author{Patrick~Murphy, 
        Ashutosh~Sabharwal 
        and~Behnaam~Aazhang
\\
Department of Electrical and Computer Engineering, \\
Rice University, \\ Houston, TX, 77005
\thanks{This work was partially funded by NSF grants CNS-0325971, CNS-0619767 and CNS-0551692.}%
\thanks{Manuscript received July 1, 2007.}}


\maketitle

\begin{abstract}
In this paper, we present the results from over-the-air experiments of a complete
implementation of an amplify and forward cooperative communications system. Our custom OFDM-based physical layer uses a distributed version of the Alamouti block code, where the relay sends one branch of Alamouti encoded symbols. First we show analytically and experimentally that amplify and forward protocols are unaffected by carrier frequency offsets at the relay. This result allows us to use a conventional Alamouti receiver without change for the distributed relay system. Our full system implementation shows gains up to 5.5dB in peak power constrained networks.  Thus, we can conclusively state that even the simplest form of relaying can lead to significant gains in practical implementations.
\end{abstract}

\section{Introduction}

Cooperative communications~\cite[and references therein]{Sendonaris:2003fk,Nosratinia:2004uq} has emerged as a significant concept to improve reliability and throughput in wireless systems. In cooperative communications, the resources of distributed nodes are effectively pooled for the collective benefit of all nodes. While cooperation can occur at different network layers (and hence at different time scales), physical layer cooperation at symbol time scales offers the largest benefit. However, symbol level cooperation is also potentially hardest to implement due to significant challenges in enabling it in distributed systems. In this paper, we take the first significant steps in building and understanding the issues in implementing practical cooperative communication systems.

We focus our attention on amplify and forward protocols~\cite{Laneman:2003lr} where the relay node simply amplifies and retransmits the analog waveform received from the source node. This simple protocol was shown to increase the diversity order~\cite{Laneman:2003lr}, allowing single-antenna nodes to cooperate and achieve performance like a real MIMO system. However, most analyses to date have ignored the challenge of implementing such a distributed space-time scheme in the face of analog and digital distortions like carrier frequency offset, inaccurate synchronization and gain control for analog to digital conversion. All of these are significant parts of practical wireless systems which, if handled poorly, can cause significant performance degradation.\footnote{Note that additive white Gaussian noise model with multiplicative fading is a highly oversimplified abstraction of an actual wireless link, which almost always has to deal with severe nonlinearities, finite precision and lack of shared clock references.}

Our contributions are three-fold. First, we show analytically and experimentally that amplify and forward protocols are not affected by the carrier frequency offset of the relaying nodes. That is, the final received signal at the destination is only affected by the carrier offset between the source and destination, much like \emph{relay-less} system. This is significant finding which shows that from the point of view of the destination, it can use a receiver built for a conventional multiple antenna transmissions without employing a multiuser-like front-end to handle non-coherent transmissions from multiple nodes.

The above finding leads to the second contribution which allows us to use a traditional Alamouti receiver \emph{without} any change for the relay system. In fact, the destination can be potentially made agnostic of the fact whether the transmission is 1$\times$1 (SISO, one transmit antenna completely off), 2$\times$1 (MISO Alamouti) or 1$\times$1$\times$1 (relay system with distributed Alamouti).

Lastly, we build a fully operational amplify and forward system which assumes only time synchronization between source and relay to mimic packet synchronous systems like GSM or WiMAX. The system is built using the resources of the Rice University Wireless Open Access Research Platform (WARP)~\cite{warpURL} and implements a high-speed wireless link using an Alamouti-encoded OFDM physical layer. With a peak power constraint per node, relaying adds more power to the system leading to gains up to 5.5dB in BER performance for both BPSK and QPSK systems in \emph{actual} wireless channels. With a total power constraint, where the total transmit power of the relaying system is same as that of the point-to-point system, the relaying systems gains are still 2dB or more. The gains can be attributed to a mix of diversity benefits and reduction in effective path loss due to relay location.

We immediately note that our work has only scratched the surface in exploring the issues in implementing cooperative systems. For example, we have only partially optimized the parameters in the receiver front-end (e.g. automatic gain control) and the choice of amplify and forward schemes. Despite of these suboptimal elements, we show that cooperation can still lead to significant gains in real implementations with commercial grade components. As obvious extensions to this work, we will implement other forms of cooperation (decode and forward variants), study performance under different channel conditions and network topologies, and gain a deeper understanding in energy-performance-complexity tradeoffs.

The rest of the paper is organized as follows. In Section~\ref{sec:amplifyAndForward}, we review the amplify and forward scheme and show how it is unaffected by relay carrier frequency offset. Section~\ref{sec:ourSystem} describes our complete implementation, experimental setup and main results. We conclude in Section~\ref{sec:conclusion}.

\section{Amplify and Forward}
\label{sec:amplifyAndForward}

\begin{figure}[h]
\centering
\includegraphics[width=3.5in]{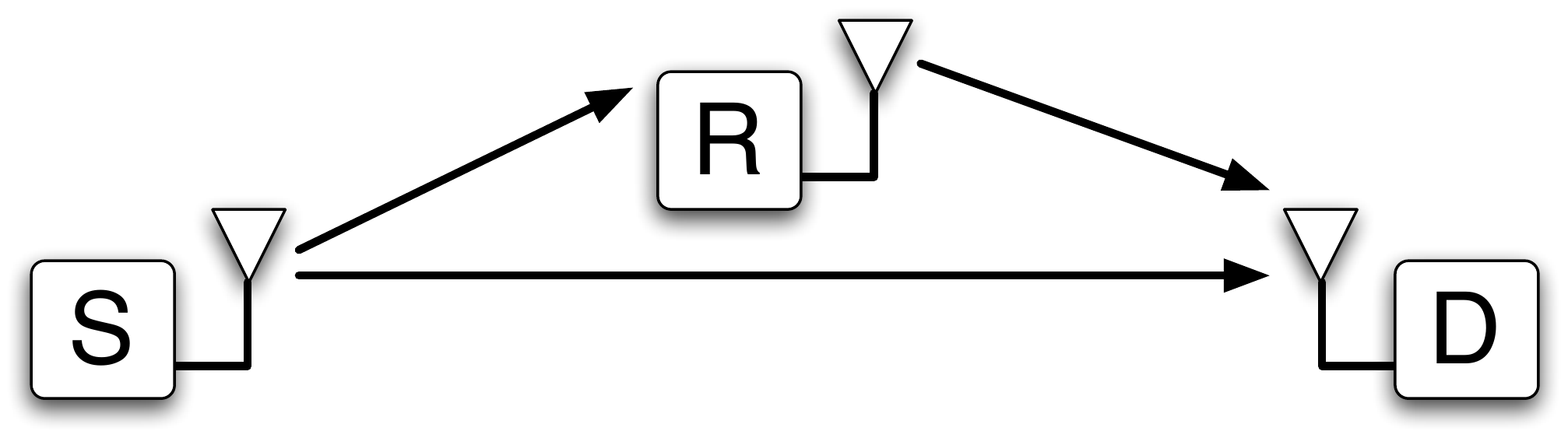}
\caption{Nodes in basic amplify and forward system}
\label{fig:nodesBasic}
\end{figure}

Amplify and forward is the simplest class of cooperative communications schemes~\cite{herhold-performance}. In amplify and forward systems, one node (the source) sends information to another (the destination). A third node (the relay) captures part of the source's transmission, amplifies it and re-transmits it without any further processing. The destination uses the combination of the source and relay's transmissions to decode the data, hopefully with fewer errors than if the source had transmitted alone. Fig.~\ref{fig:nodesBasic} shows the basic configuration of these three nodes.

The underlying idea of amplify and forward can be applied in a wide variety of ways. Our goal is to construct a cooperative system based on one realization of amplify and forward, then to use this system to explore some of the issues which arise when building a cooperative communications system.  We uncovered one particularly interesting property of amplify and forward systems, which we discuss in detail below.

\subsection{Carrier Frequency Offset}
\label{sec:cfo}

In practice, wireless nodes generate radio-frequency carriers using phase-locked loops driven by a local frequency reference. The frequency of the generated carrier varies with the frequency of the local reference. When multiple nodes use independent local references, their RF carriers differ in frequency. In most hardware, this carrier frequency offset is large enough that it must be addressed by the wireless physical layer algorithms.

Carrier frequency offset is a well-studied problem; practical algorithms exist to mitigate CFO in a wide variety of wireless systems. However, the effects of CFO have largely been ignored in the development of cooperative communication algorithms. Some schemes have been proposed which attempt to synchronize the carriers of multiple transmitting nodes in hopes their signals will constructively combine at the destination~\cite{Tu:2002yq,Brown:2005vn}. These schemes rely on some kind of shared information among transmitting nodes, either in the form of communicated phase offsets or reception of a common beacon signal. In either case, the complexity of maintaining synchronization is non-trivial.

\subsection{Radio Transceiver Model}
Our first contribution in this paper is to explore the construction of an amplify and forward system which exploits a useful property of common radio hardware. We will show how this property allows the destination node to simply ignore the carrier offset of an amplify and forward relay node.

The following analysis intentionally ignores many practical aspects of a wireless communications system, including physical layer waveform design, gains, filters, analog/digital conversion and channel effects. The goal of this derivation is solely to demonstrate the effects of carrier frequency offset in an amplify and forward link. In real communications systems, CFO is an analog (i.e. continuous time) problem, inherent in the local generation of RF carriers at each node. Thus, to trace the impact of CFO through a cooperative link, we consider only the analog baseband and RF signals in the following. The effects we omit here will certainly play a part in constructing an actual cooperative link (as described in Section~\ref{sec:ourSystem}). However, in scenarios with little Doppler effect, carrier frequency offsets can be analyzed independent of these other impairments.

\begin{figure}[h]%
\centering 
\subfloat[Transmitter model]{\includegraphics[width=2.5in]{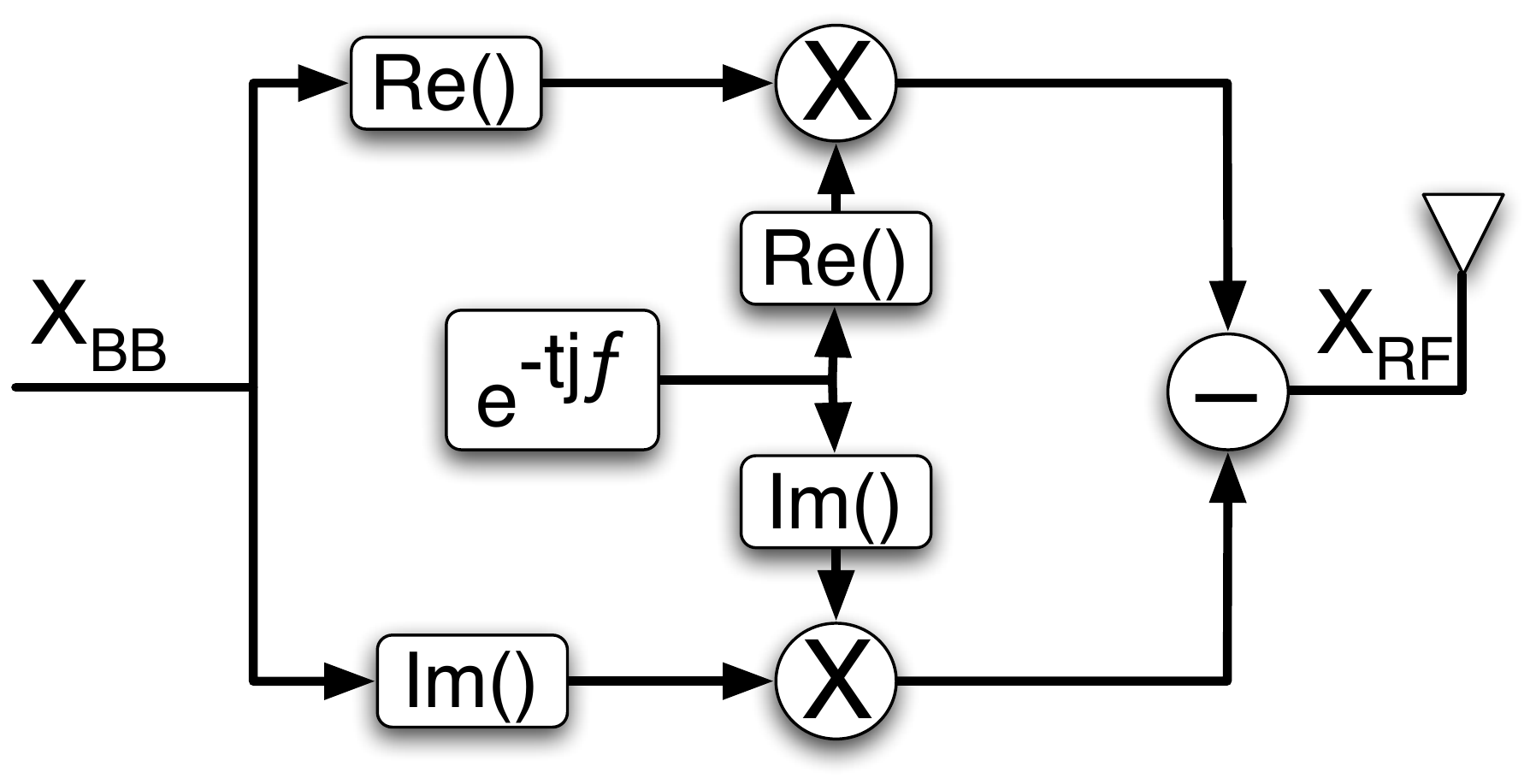}}\qquad 
\subfloat[Receiver model]{\includegraphics[width=2.5in]{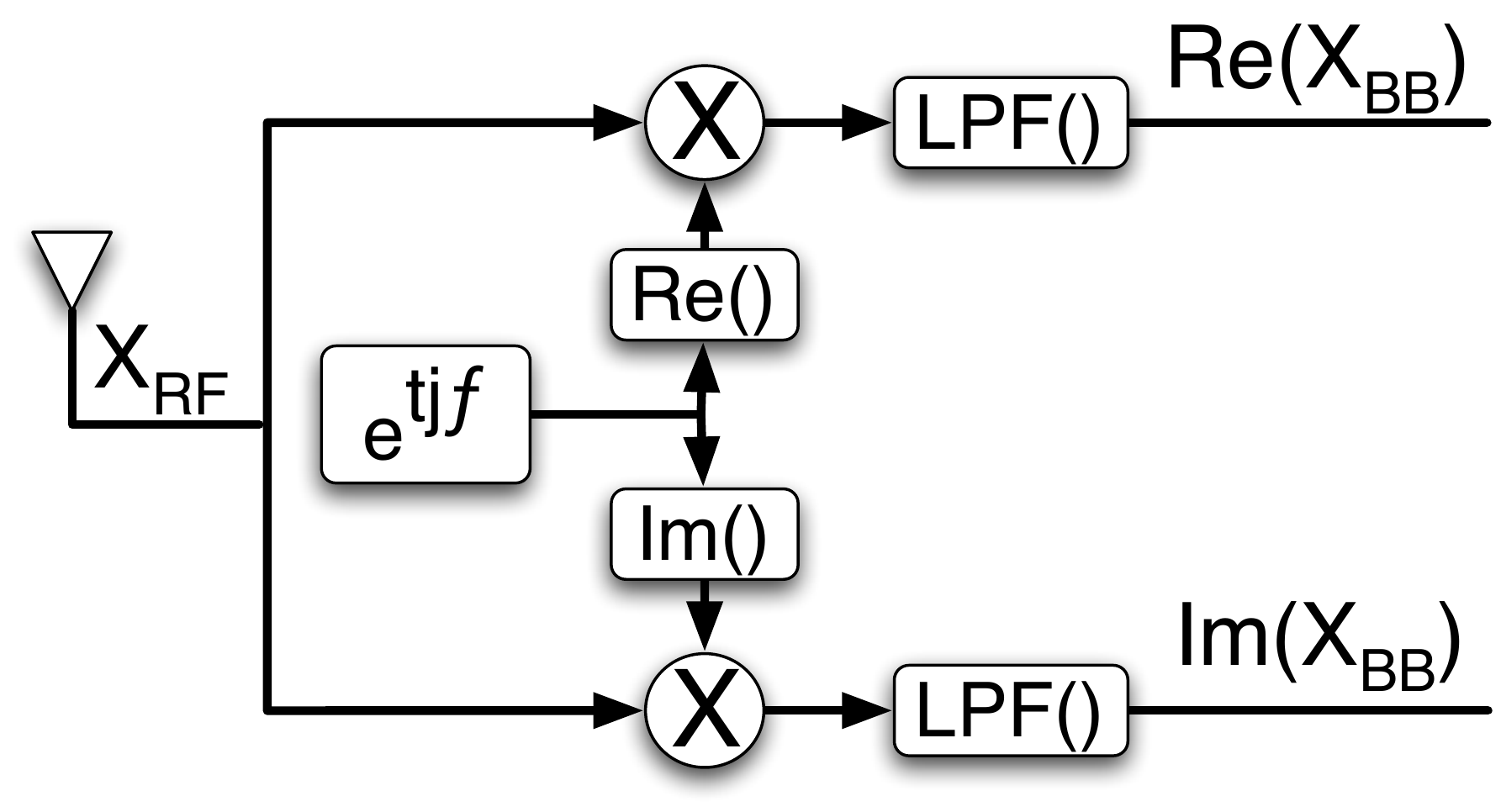}}\\ 
\caption{Tx and Rx models} 
\label{fig:txrx_blockDiagrams} 
\end{figure}

Fig.~\ref{fig:txrx_blockDiagrams} illustrates our models for the analog processes of RF upconversion and downconversion. These models reflect the inner workings of a direct conversion RF transceiver, where a common sinusoidal carrier is used for both the transmit and receive chains. The use of a common carrier reference for the transmit and receive paths at the relay node is a critical (but thankfully realistic) assumption in this analysis.

In the following, let $f_c$ denote the frequency of the carrier and $\mathsf{LPF}(x)$ a low-pass filter. Note that the baseband signals ($X_{BB}$ below) are complex, but the RF signals ($X_{RF}$ below) are all-real. This matches the implementation of wireless systems, where an RF signal is a single voltage and complex baseband signals are represented with separate $I$ and $Q$ voltages.

First, we will define the $\mathsf{Tx}()$ and $\mathsf{Rx}()$ functions which express the processes illustrated in Fig.~\ref{fig:txrx_blockDiagrams}:

\begin{equation*}
\begin{split}
X_{RF} 
	& = \mathsf{Tx}(X_{BB}, f_{C})\\
	& = \mathsf{Re}(X_{BB}) \cos(tf_{C}) - \mathsf{Im}(X_{BB}) \sin(tf_{C})\\
	& = \frac{X_{BB}e^{jtf_{C}} + X_{BB}^*e^{-jtf_{C}}}{2}\\
X_{BB}
	&  = \mathsf{Rx}(X_{RF}, f_{C})\\
	&  = \mathsf{LPF}(X_{RF}e^{jtf_{C}})
\end{split}
\end{equation*}

\subsection{CFO in Amplify and Forward}

\begin{figure}[h]%
\centering
\includegraphics[width=5.0in]{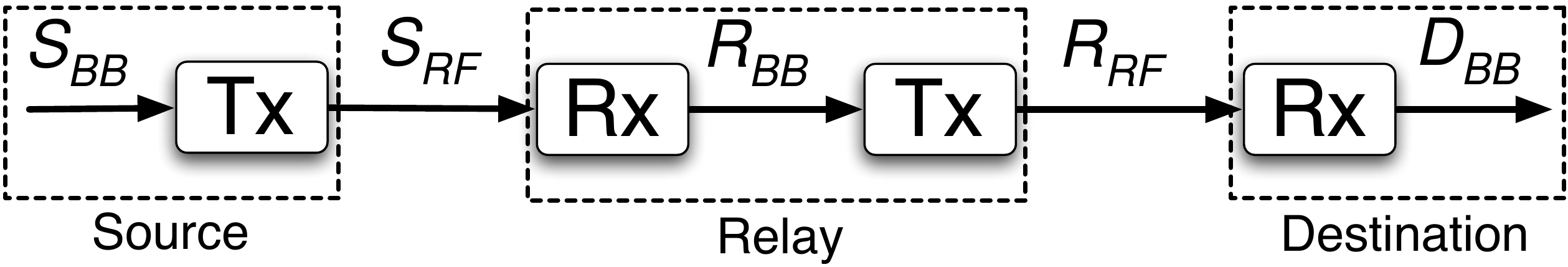}
\caption{Nodes and signals in an amplify and forward link}
\label{fig:nodesSignals}
\end{figure}

We will now apply these functions to trace the frequency offset of a signal as it propagates through an amplify and forward cooperative link. Fig.~\ref{fig:nodesSignals} illustrates the nodes and signal names used in the following derivation.

Consider the initial source to relay transmission, where $f_{CS}$ and $f_{CR}$ are the carrier frequencies of the source and relay, respectively. The sequence of operations and corresponding signals are:

\begin{equation*}
S_{BB} \rightarrow \mathsf{Tx}(S_{BB}, f_{CR}) \rightarrow S_{RF} \rightarrow \mathsf{Rx}(S_{RF}, f_{CR}) \rightarrow R_{BB}
\end{equation*}
which when expanded gives the following, assuming $\mathsf{LPF()}$ is a linear filter with gain $2$:
\begin{equation}
\begin{split}
S_{RF} 
	& = \mathsf{Tx}(S_{BB}, f_{CS})\\
	& = \frac{S_{BB}e^{jtf_{CS}} + S_{BB}^*e^{-jtf_{CS}}}{2}\\
R_{BB}
	& = \mathsf{Rx}(S_{RF}, f_{CR})\\
	& = \mathsf{LPF}(S_{RF}e^{jtf_{CR}})\\
	& = \mathsf{LPF}\left(
		\frac
		{(S_{BB}e^{jtf_{CS}} + S_{BB}^*e^{-jtf_{CS}})(e^{jtf_{CR}})}
		{2}
		\right)\\
	& = \mathsf{LPF}\left(
		\frac
		{S_{BB}e^{jt(f_{CS}-f_{CR})} + S_{BB}^*e^{-jt(f_{CS}+f_{CR})}}
		{2}
		\right)\\
	& = S_{BB} (e^{jt(f_{CS}-f_{CR})}).
\end{split}
\end{equation}
As expected, the baseband signal received at the relay suffers a frequency offset due to the difference between the source and relay carrier frequencies.

Next, we trace the transmission from relay to destination:

\begin{equation*}
R_{BB} \rightarrow \mathsf{Tx}(R_{BB}, f_{CR}) \rightarrow R_{RF} \rightarrow \mathsf{Rx}(R_{RF}, f_{CD}) \rightarrow D_{BB}
\end{equation*}

\begin{equation}
\begin{split}
R_{RF} 
	& = \mathsf{Tx}(R_{BB}, f_{CR})\\
	& = \frac{R_{BB}e^{jtf_{CR}} + R_{BB}^*e^{-jtf_{CR}}}{2}\\
D_{BB}
	& = \mathsf{Rx}(R_{RF}, f_{CD})\\
	& = \mathsf{LPF}(R_{RF}e^{jtf_{CD}})\\
	& = \mathsf{LPF}\left(
		\frac
		{(R_{BB}e^{jtf_{CR}} + R_{BB}^*e^{-jtf_{CR}})(e^{jtf_{CD}})}
		{2}
		\right)\\
	& = \mathsf{LPF}\left(
		\frac
		{R_{BB} e^{jt(f_{CR}-f_{CD})} + R_{BB}^* e^{-jt(f_{CR}+f_{CD})}}
		{2}
		\right)\\
	& = R_{BB} (e^{jt(f_{CR}-f_{CD})}).
\end{split}
\end{equation}
Finally, we substitute the previous expression for $R_{BB}$:
\begin{equation}
\begin{split}
D_{BB}
	& = (S_{BB} e^{jt(f_{CS}-f_{CR})})(e^{jt(f_{CR}-f_{CD})})\\
	& = S_{BB} (e^{jt(f_{CS}-f_{CD})}).
\end{split}
\end{equation}
Thus, the received baseband signal at the destination node suffers a frequency offset determined solely by the difference between $f_{CS}$ and $f_{CD}$, independent of their respective offsets from $f_{CR}$. In other words, the relay's carrier frequency offset with respect to the source and destination nodes does not affect the final signal received at the destination.

\subsection{Empirical Verification}
\label{sec:cfoDemo}
In order to substantiate the preceding analysis and to verify the impact of its inherent assumptions, we constructed an RF link which allows the direct observation of carrier frequency offsets. In this setup, one node acts as both the source and destination, while a second node acts as the relay. The source generates a constant valued baseband signal, which after upconversion results in the transmission of a sinusoid at exactly $f_{CS}$ (i.e. intentional carrier leakage). The relay node receives this sinusoid, downconverts it with its local carrier and saves the samples at baseband. If the analysis is correct, these samples should be of a sinusoidal signal with frequency $(f_{CS} - f_{CR})$. The relay then transmits the same samples back to the first node. If our assumptions and analysis hold, the first node should receive a constant valued signal at baseband, showing no frequency offset as a result of amplification and retransmission at the relay.

\begin{figure}[h]%
\centering 
\subfloat[$\approx$10ms Delay]{\label{fig:cfoDemo_short} \includegraphics[width=3in]{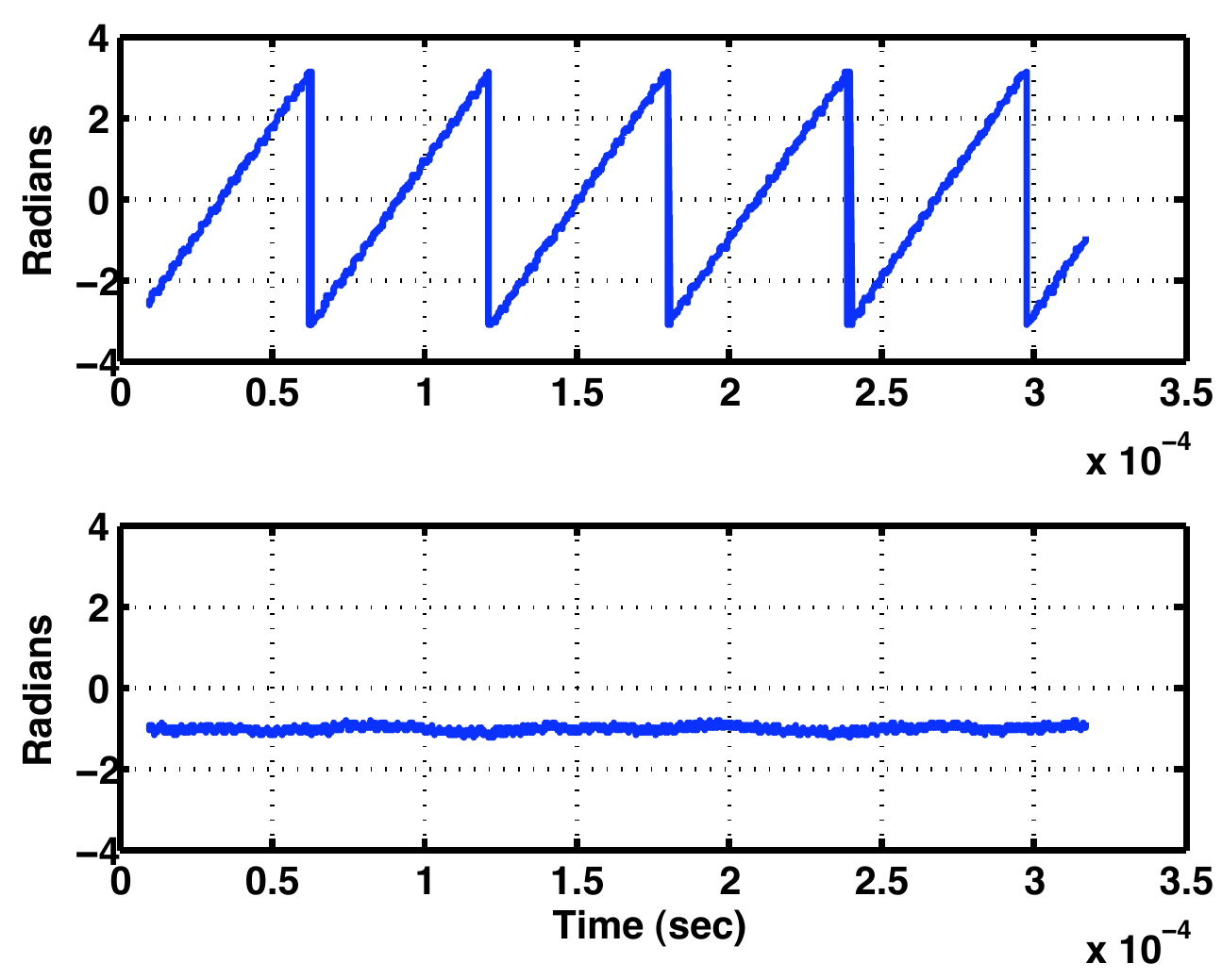}}\qquad 
\subfloat[$\approx$2min Delay]{\label{fig:cfoDemo_long} \includegraphics[width=3in]{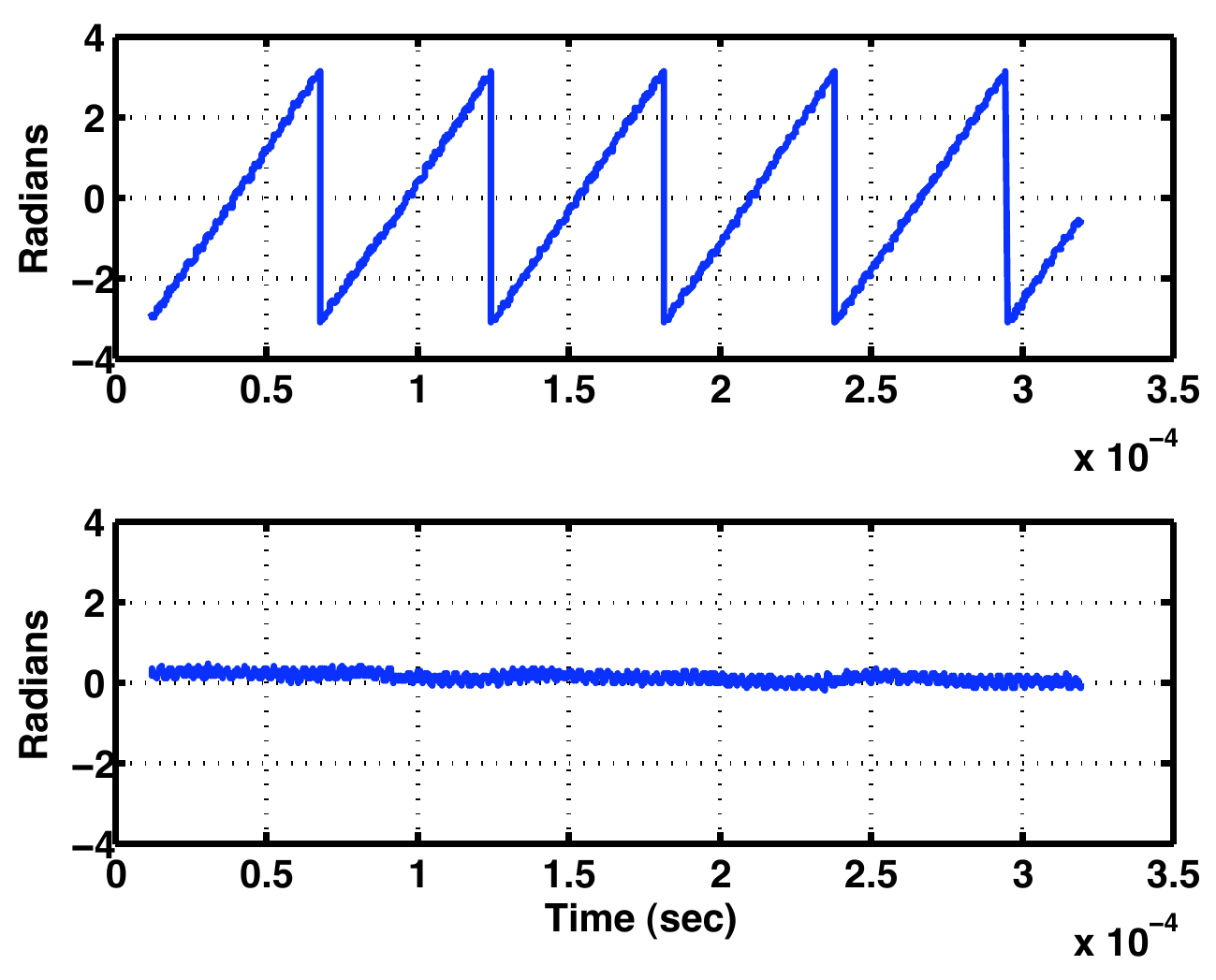}}\\ 
\caption{Experimental observation of carrier offset in a relay system} 
\label{fig:cfoDemot} 
\end{figure}

Fig.~\ref{fig:txrx_timing} shows the results of this experiment. Two trials are depicted here. In the first, the relay transmits its received signal after a short delay, approximately 10msec. In the second, the relay waits two minutes before re-transmitting. The transmission in both directions happens over a wire to eliminate any channel effects. The top plots depict the phase of the signal received at the relay. The phase of this signal is increasing linearly in time, corresponding to a received sinusoid. This sinusoid is the direct result of carrier frequency offset between the two nodes. The bottom plots depict the phase of the signal received at the source node, after it is buffered and re-transmitted by the relay. The complete lack of the saw wave pattern clearly illustrates the relay canceling its own carrier offset during re-transmission. In Fig.~\ref{fig:cfoDemot}(b), a very slight slope can be observed in the received signal's phase. This is the result of a minor drift in the node's local oscillator frequency. The WARP hardware utilized in this experiment uses temperature-compensated crystal oscillators for the carrier reference, which accounts for the very minor drift, even after two minutes. Cheaper oscillators, like those used in low-end commercial wireless hardware, could exhibit larger drifts over time.

\section{Building a Cooperative System}
\label{sec:ourSystem}
This section describes the construction of an amplify and forward cooperative communications system which relies on the properties described in Section~\ref{sec:amplifyAndForward}. This system is implemented on WARP~\cite{warpURL}, making heavy use of the custom hardware, physical layer designs and other support packages provided by the platform.

\subsection{Overview}
\begin{figure}[h]
\centering
\includegraphics[width=3.5in]{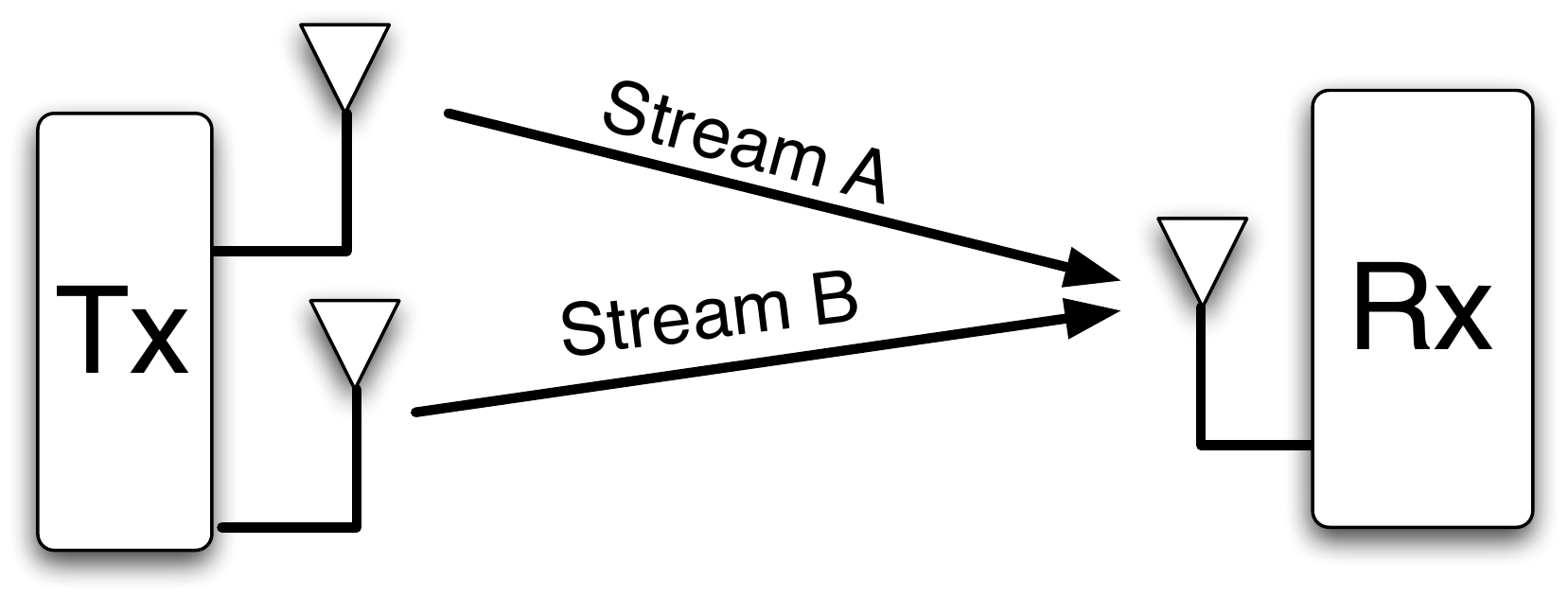}
\caption{Equivalent 2$\times$1 Alamouti setup}
\label{fig:2x1AlamoutiSetup}
\end{figure}

Our system is built on the idea of distributed space time coding~\cite{Anghel:2003fj,Laneman:2003lr}, where multiple nodes cooperate to transmit a signal which approximates the transmission of a single, multiple-antenna node. In particular, we employ Alamouti's space time block code (STBC)~\cite{Alamouti:1998kx}. Fig.~\ref{fig:2x1AlamoutiSetup} illustrates the classic 2$\times$1 STBC configuration which the proposed cooperative scheme imitates. The signal names here correspond to the two spatial streams generated by a two-antenna Alamouti transmitter; these signals play a key role the proposed cooperative version of this link.

The Alamouti STBC encodes two data symbols across two symbol periods and two spatial streams. Given two data symbols $x_0$ and $x_1$, the code outputs the signals shown in Table~\ref{tab:alamouti}. In each symbol period at the receiver, the superposition of the two streams is received after each passes through separate channels; the signals received in two symbol periods are represented by $r_0$ and $r_1$ below. The receiver uses local channel estimates and the following combining rules to recover the original data symbols:

\begin{equation}
\label{eq:alamoutiCombining}
\begin{split}
	r_0
		& = h_Ax_0 + h_Bx_1 + n_0\\
	r_1 
		& = -h_Ax_1^* + h_Bx_0^* + n_1\\
	\tilde{x}_0
		& = h_A^*r_0 + h_B r_1^*\\
	\tilde{x}_1
		& = h_B^*r_0 - h_A r_1^*.
\end{split}
\end{equation}

\begin{table}[h]
\caption{Alamouti STBC encoding}
\label{tab:alamouti}
\centering
\begin{tabular}{|c||c|c|}
\hline
 & $t_0$ & $t_1$\\
\hline
Stream A & $x_0$ & $-x_1^*$\\
\hline
Stream B & $x_1$ & $x_0^*$\\
\hline
\end{tabular}
\end{table}

\begin{figure}[h]%
\centering 
\subfloat[Time slot 1]{\includegraphics[width=3in]{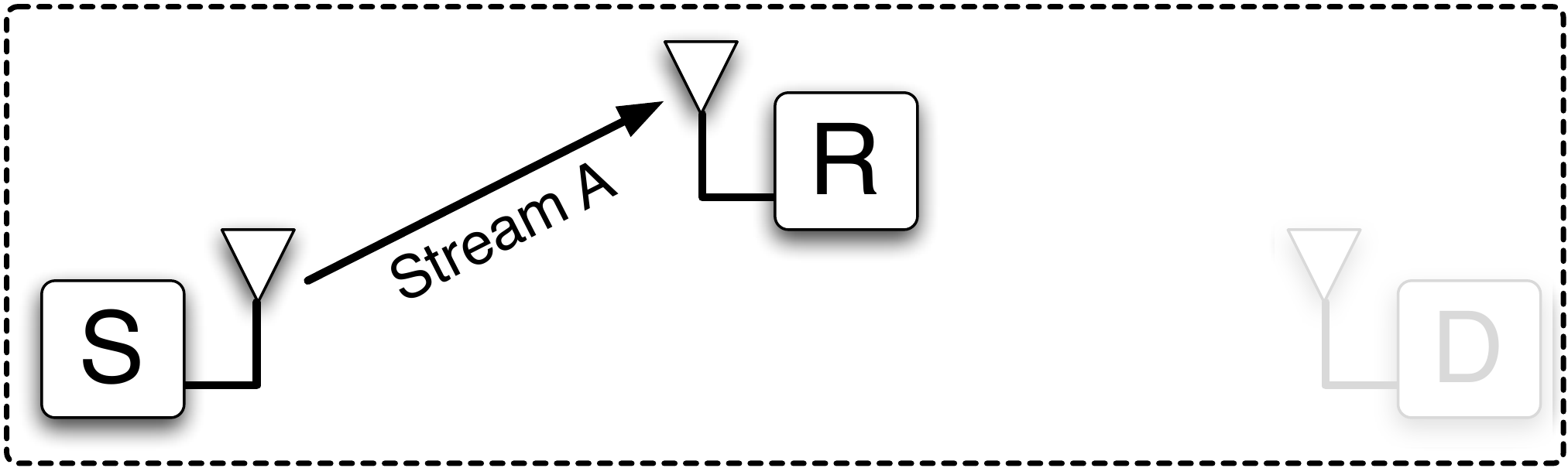}}\qquad 
\subfloat[Time slot 2]{\includegraphics[width=3in]{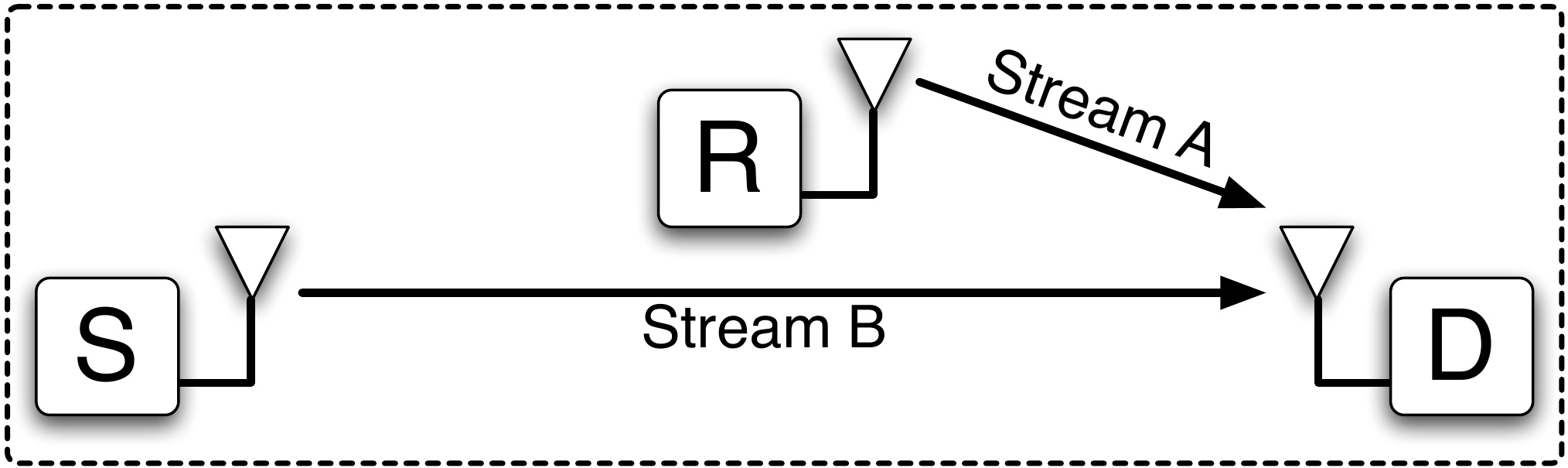}}\\ 
\caption{Node configurations and activity in the amplify and forward system} 
\label{fig:txrx_timing} 
\end{figure}

Much like other cooperative protocols for half-duplex radios, the proposed cooperative link operates in two time slots per packet. Fig.~\ref{fig:txrx_timing} illustrates the activity of each node in our scheme's two time slots. In the first slot, the source node transmits the full packet, encoded using the Alamouti space-time block code. This signal matches that which would be sent from one antenna in a true two-antenna Alamouti transmission. The relay node receives this transmission and stores the raw samples in a buffer. In the second time slot, the source node transmits the other half of the Alamouti-encoded sequence, and the relay transmits its stored copy of the first transmission. The destination node receives the superposition of these simultaneous transmissions. From the perspective of the destination, it receives a standard Alamouti-encoded packet, where each of the spatial components was exposed to an independent channel.

\subsection{Physical Layer Design}

In the proposed amplify and forward scheme, the timing of the two transmissions in the second time slot cannot be perfectly guaranteed. The offset between the arrival times of the source and relay's transmissions can be modeled as multipath. This is analogous to the signals sent from a standard two antenna Alamouti transmitter arriving at slightly different times at the receiver after passing through different channels.

In order to cleanly handle this potential impairment, we chose OFDM as the underlying physical layer for our cooperative system. OFDM's inherent immunity to multipath makes it an ideal PHY for an amplify and forward system, as a delayed transmission is treated as just another reflection in the channel.

The details of the physical layer design are described below.

\subsubsection{Frame Format}
Our cooperative physical layer uses the following frame format, partially inspired by IEEE 802.11a~\cite{IEEE80211a}. The transmissions are composed of four components:

\begin{itemize}
	\item{Short training symbols (STS): 10 16-sample sequences, used for AGC convergence}
	\item{Long training symbols (LTS): 2.5 64-sample sequences, used for fine symbol timing}
	\item{Channel training symbols (Tr$_A$/Tr$_B$): 80-sample channel training symbols}
	\item{Spatial streams: Payload data encoded with Alamouti's space-time code}
\end{itemize}

\begin{figure}[h]
\centering
\includegraphics[width=6.5in]{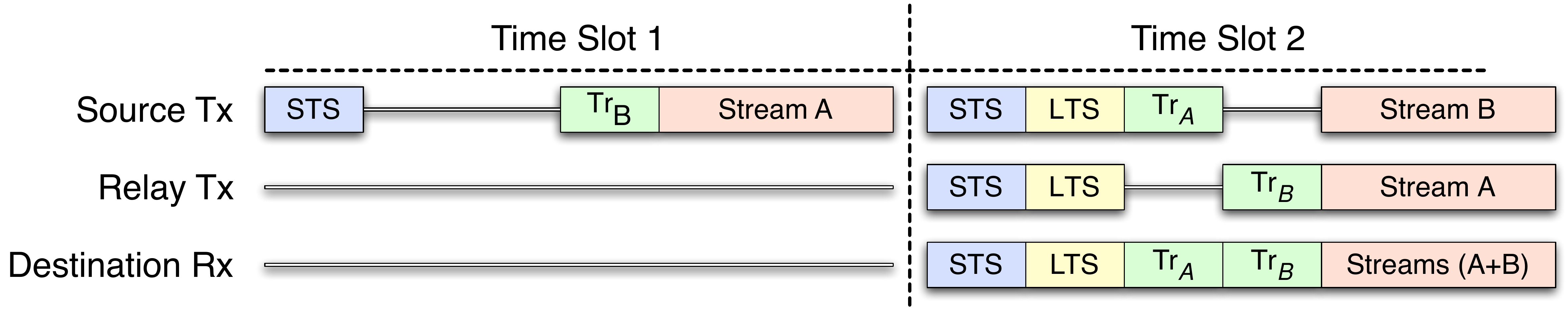}
\caption{Physical layer frame formats}
\label{fig:frameFormats}
\end{figure}

The composition of the transmissions in each time slot are illustrated in Fig.~\ref{fig:frameFormats}.

In the first time slot, the source node transmits a frame designed to trigger packet detection at the relay but avoid packet detection at the destination. This is achieved by omitting the long training symbols. Our OFDM receiver uses a correlator to search for the LTS within a fixed window after an energy detection event. If this correlation fails, the receiver assumes a false packet detection and resets. The relay node does not perform this check.

In the second time slot, the LTS must be included in the transmissions to allow the destination to properly detect the packet and synchronize the receiver. The source node transmits the full STS/LTS preamble in this slot. The relay node also sends a full preamble in the second time slot. The relay stores this preamble in a lookup table and sends it in place of the STS captured in the first slot. After sending the LTS, the relay begins transmitting the buffered samples it captured, offset by the proper amount to keep the two transmissions aligned in time.

We designed this scheme to allow successful packet detection and synchronization at the destination even if it receives just one of the two transmissions in the second time slot. In an intuitive sense, this scheme preserves full diversity as it will fail only if two (presumably) independent channels are simultaneously in deep fades.

\subsubsection{Synchronization}
In our setup, the relay node uses a dedicated synchronization signal from the source to initiate its buffering and re-transmission processes. The packet lengths are also fixed throughout our experiments and are known ahead of time by every node. This kind of synchronization mimics what is used in scheduled access systems like GSM or WiMAX.

The destination node implements autonomous packet detection. This system uses the RSSI (received signal strength indicator) signal from the RF transceiver to detect a spike in received energy indicating a the start of a new packet. The timing of the packet is refined in the PHY by cross-correlation against the LTS in the packet's preamble. This is the same approach to packet detection and timing used in a non-cooperative random access system. If the uncertainty of packet arrival times at the destination were eliminated, as in slotted systems like GSM or WiMAX, we expect the system performance would improve.

Every node has independent sampling and radio reference clocks. Given the relatively short packets, we ignore sampling frequency offsets throughout. Offsets among the radio reference clocks result in carrier frequency offsets, the effects of which we explored in Section~\ref{sec:cfo}.

\subsubsection{Gain Control}
Both the relay and destination nodes implement automatic gain control, which executes with each packet detection. The AGC algorithm sets the gains for the receive amplifiers in the RF transceiver in the first 2-3$\mu$sec after packet detection, well within the STS section of the preamble.

The relay node amplifies its received signal in both the analog and digital domains. The relay's RF transceiver uses low-noise amplifiers to boost the analog RF and baseband signals in the receive path. The gain settings for these amplifiers are chosen for each packet by the AGC system. The result of this amplification is an analog signal whose amplitude is independent of the received power. This signal is sampled by the relay's ADC and buffered in the FPGA. During the second time slot, the relay multiplies these stored samples by a constant before driving them into the DAC. The radio board's RF transceiver and power amplifier apply the final stages of gain before transmission. The digital gain value is fixed, as it is determined solely by the difference in the ADC and DAC dynamic ranges and does not depend on the RF transceiver's gain settings.

\subsubsection{Channel Estimation}
The destination must estimate two channels in order to properly combine the Alamouti-encoded symbols, analogous to the two channels in a classic 2$\times$1 Alamouti configuration. In our setup, however, one of these channels is actually the combination of two physical channels: source-to-relay and relay-to-destination. Only the relay's retransmitted signal experiences this compound channel. The destination node uses a training symbol originally embedded by the source, then retransmitted by the relay, to estimate the compound channel. The second channel the destination must estimate is the source-destination channel using a training symbol embedded in the source's transmission in the second time slot. The source node constructs its transmissions so that in the second time slot, the two training symbols do not overlap, allowing independent estimates at the destination node.

\subsubsection{OFDM}
The source and destination nodes implement identical, full Alamouti OFDM transceivers. This PHY was originally implemented for use in a standard 2$\times$1 Alamouti OFDM link. Due to the structure of our amplify and forward configuration, the same receiver design works as-is, without modification, in the cooperative system. The transmitter design requires minor modifications to enable the back-to-back transmissions of the spatial streams from a single antenna. The universality of the receiver design which functions without modification in 1$\times$1, 2$\times$1 and 1$\times$1$\times$1 configurations is a significant benefit of amplify and forward systems.

All processing in the PHY is implemented in the WARP FPGA and executes in real-time. Carrier frequency offset estimation, symbol timing estimation, phase noise tracking, equalization and detection are all implemented in fixed-point in the FPGA. The physical layer operates in a 12.5MHz bandwidth with a raw data rate of 7.5 or 15Mbps by transmitting BPSK or QPSK symbols in 52 of 64 subcarriers. One training symbol is used per channel, and 4 pilot subcarriers are used to track phase noise and residual carrier offset.

\subsection{Experiment Design}

Our experiments were conducted in a three node setup, each implementing a single antenna half-duplex transceiver. The nodes were built with WARP hardware, with one FPGA board and one radio board~\cite{warpURL}. The physical locations of the nodes and their antennas were fixed throughout the experiment. The relay's antenna was approximately half-way between the source and destination node. All transmissions were over-the-air in the 2.4GHz ISM band. A channel not used by any other wireless devices was chosen to minimize the impact of interference on our BER results. All experiments were conducted indoors. As a result, there was very little mobility in the channel during each experiment and each node had a clear line-of-sight to the other two.

Given the nodes' locations are fixed throughout, we used the transmission power of the source and relay nodes as a proxy for SNR and as the independent variables in the results below.

The transmit power and received gains are adjusted inside the WARP radio board's RF transceiver. The various gain stages are applied to the analog and RF signals by low-noise amplifiers. The analog signals at the ADCs and DACs are always the same amplitude, so the contribution of quantization to the overall performance is fixed and independent of a node's transmit or receive power.

\subsection{Results}
\begin{figure}[h!]
\centering
\includegraphics[width=5.0in]{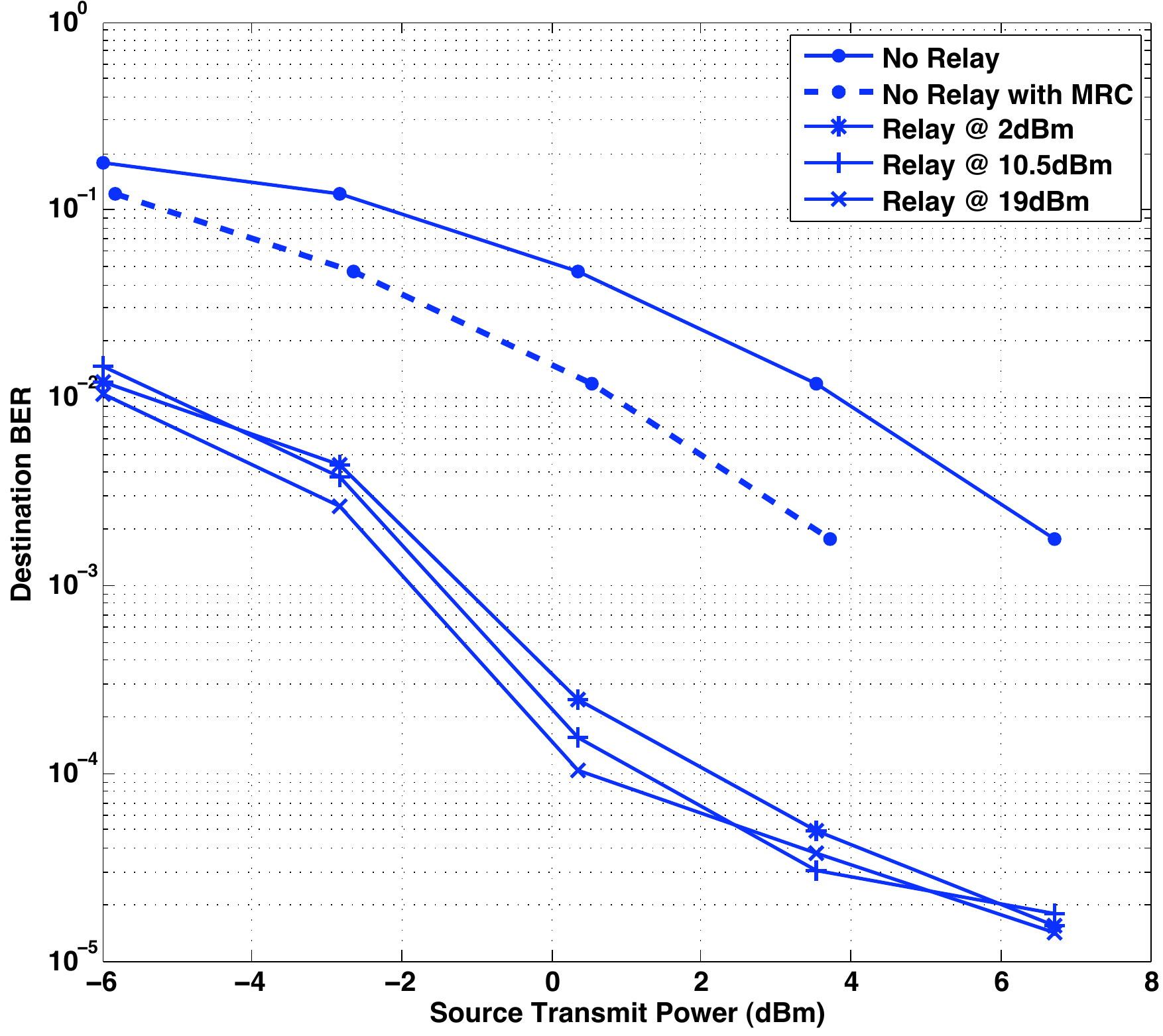}
\caption{BER results for relay vs. relay-less links}
\label{fig:berResults_qpsk}
\end{figure}

From the plots in Fig.~\ref{fig:berResults_qpsk}, it is immediately clear that the relay node significantly improves the BER performance in the cooperative link. 

The top curve shows the performance of the non-cooperative link. For these tests, the source and destination nodes operate exactly as described above, but the relay is switched off. A copy of this curve shifted left 3dB is also included. This shifted curve illustrates the best possible performance the destination could achieve if it performed maximal ratio combining (MRC) on the two copies of each packet it receives, instead of simply ignoring the energy it received in the first time slot.

A second observation we can make from these results is whether adding a relay helps even if the system's total transmit power were artificially constrained. To make this comparison, we first choose a point along the X-axis, determine the total transmit power (source + relay), then find the point on the X-axis of equal power. The comparison of relay-aided vs. no-relay BER values at these two points reveals whether fixed total power is better allocated to the relay or source.

\begin{figure}[h!]
\centering
\includegraphics[width=4.5in]{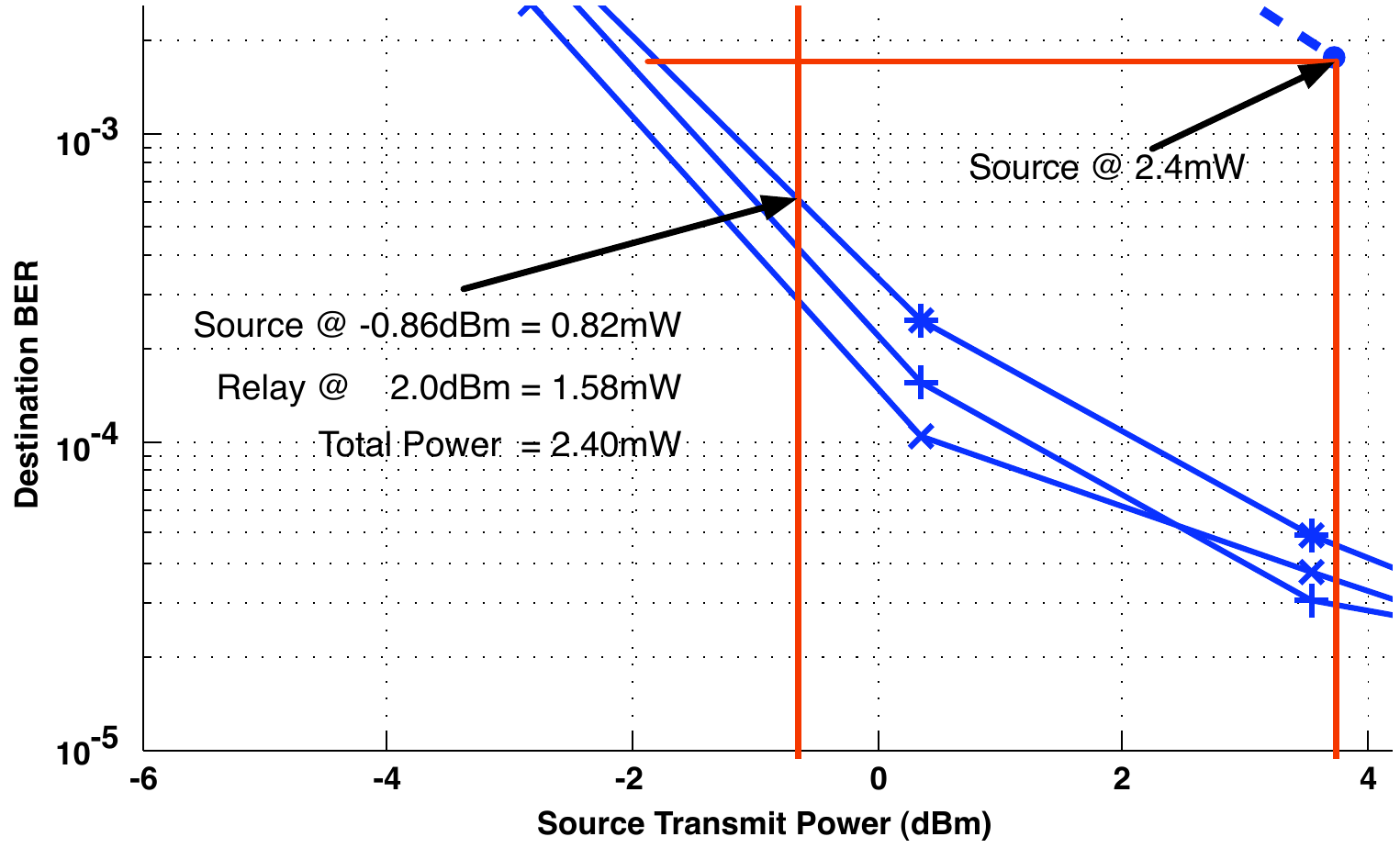}
\caption{BER results with equivalent power comparison}
\label{fig:berResults_powerZoom}
\end{figure}

Fig.~\ref{fig:berResults_powerZoom} shows a region from the BER plot in more detail and illustrates this comparison. It can be seen that allocating some power to the relay outperforms the comparable no-relay configuration by at least 5dB. This gain is heavily dependent on the network topology and channel conditions. An exhaustive study of networks would be required to state this result more generally. However, this example still clearly demonstrates that given a total power constraint, the tradeoff between source and relay transmit power can favor allocating power to the relay in some realistic situations.

A final point to observe in these results is the relatively minor performance improvement which results from extra transmission power at the relay. This strongly indicates that the source-relay link dominates the overall performance. This fits the intuitive notion that if the relay receives a poor signal in the first time slot, it will spend most of its power retransmitting noise, with little benefit to the destination.

\begin{figure}[h!]
\centering
\includegraphics[width=5.0in]{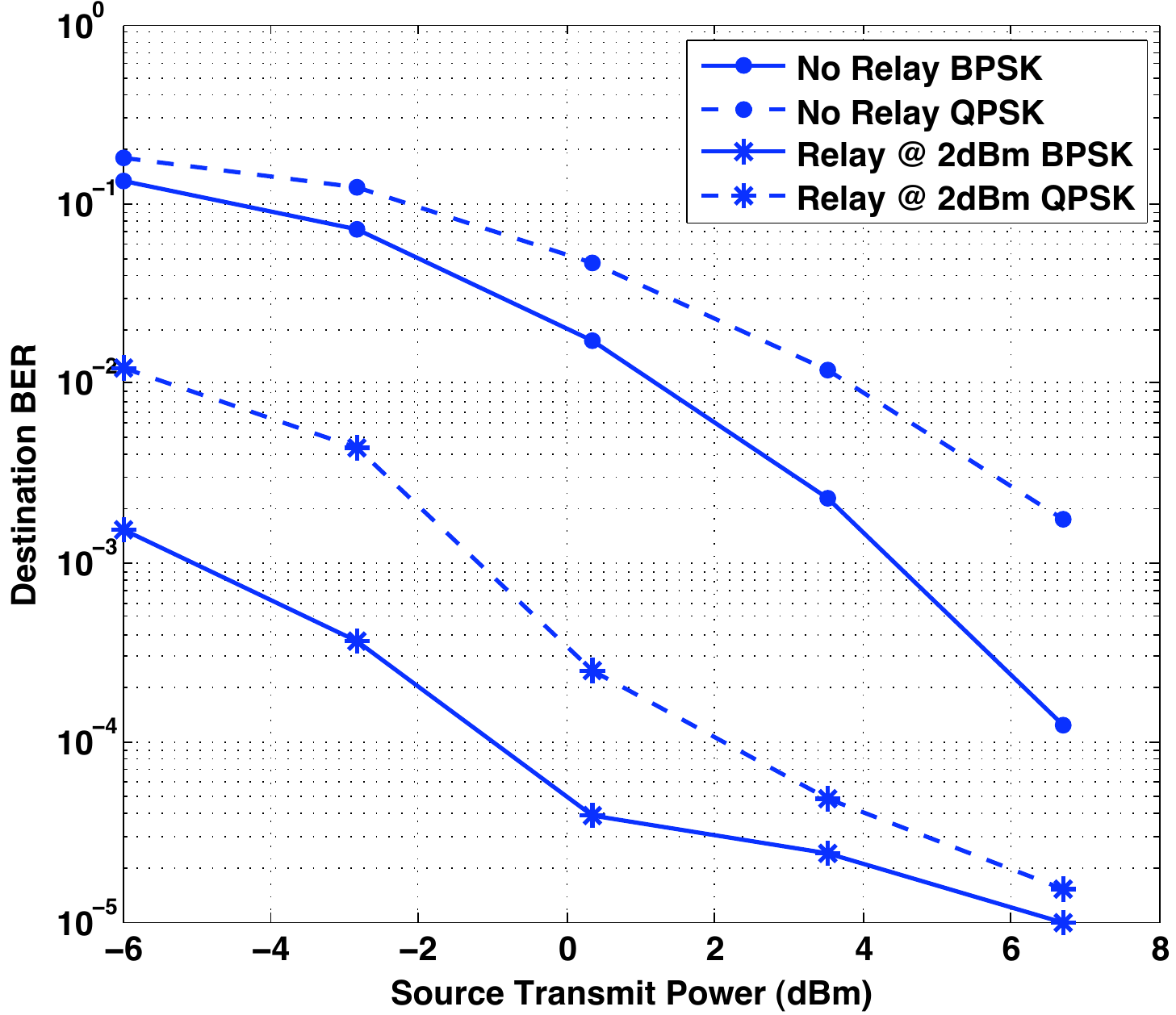}
\caption{BER results for BPSK and QPSK modulations}
\label{fig:berResults_modCompare}
\end{figure}

The second set of results in Fig.~\ref{fig:berResults_modCompare} compares the performance of two modulation schemes. For these tests, we configured the source and destination to use either BPSK or QPSK on the same 48 (of 64) subcarriers used in the previous trials. For BSPK, half as many bytes were sent per packet in order to maintain the same source/relay timing parameters used for QPSK transmissions. As expected, BPSK performs better than QPSK, and both schemes benefit when using the relay. This plot offers another interesting comparison. The total transmission duration for relay-aided QPSK and relay-less (i.e. single time slot per packet) BPSK are roughly equivalent. Thus, the source can choose between the two schemes without significantly affecting medium access control. Given this choice, the results in Fig.~\ref{fig:berResults_modCompare} indicate that at low to moderate SNR, the source should choose relay-aided QPSK over relay-less BPSK.

\subsection{Comments}

Our observation of carrier offset cancellation at the relay is based on a few important but realistic assumptions. First, the magnitude of the relay's offset must be small relative to the signal's bandwidth. If the offset is too large, the resulting baseband spectrum at the relay will be shifted into the stop band of the transceiver's low pass filters. The same problem would occur in a non-cooperative system if the source-destination CFO were too large. As long as the wireless hardware uses oscillators of sufficient accuracy to allow non-cooperative links, our CFO-free amplify and forward observation will hold. The second assumption is that the relative frequencies of the carrier frequencies at each node do not drift significantly between time slots. This is again a function of the quality of the system's oscillators. An oscillator's frequency stability over temperature and time is generally well specified by the manufacturer and can be used to determine the expected frequency drift. In practice, frequency changes on per-packet time scales are very small (as we demonstrate in Section~\ref{sec:cfoDemo} above).

OFDM is a natural PHY for an amplify and forward system, as it allows the re-transmitted signal to be treated as just another multipath reflection at the destination. However, this approach reduces the overall delay spread tolerance of the OFDM PHY, as it uses some part of each OFDM symbol's cyclic prefix to account for inaccuracies in the timing of the relay's transmission. While the size of the cyclic prefix (1.28$\mu$sec in our case) in OFDM systems is generally conservative, especially for stationary indoor environments, this loss of delay spread tolerance could impact performance in more hostile channels.

Finally, we note that our results are only a first but important step towards studying cooperative communications in practice. We observed real performance gains when using amplify and forward, but the magnitude of these gains are certainly subject to many parameters, including network topology, channel conditions and physical layer design.

\section{Conclusion}
\label{sec:conclusion}
In this work, we have built a full cooperative communications system which operates in real-time over real wireless channels. This system puts to practice some ideas from existing work in cooperative communications. It also relies on our own results in understanding carrier frequency offset in amplify and forward systems. Our performance evaluation shows a clear benefit to using amplify and forward relays, demonstrating a significant BER improvement under realistic wireless conditions. It is clear that physical layer cooperation is largely uncharted territory, especially with regard to implementation in practical systems. To enable a deeper understanding, our implementation of an amplify and forward cooperative system will be available in the Rice WARP project's open-source repository~\cite{warpURL}, allowing the community to systematically evaluate different relaying protocols over real wireless channels with all practical considerations.

\section*{Acknowledgment}

The authors would like to thank Dr. Chris Dick at Xilinx and the Xilinx University Program for their continuing support of the WARP project.

\bibliographystyle{IEEEtran}
\bibliography{bib/IEEEabrv,bib/JSAC}

\end{document}